\pdfoutput=1

\documentclass[a4paper]{article}
\usepackage{mathptmx}
\usepackage{graphicx}
\usepackage{dcolumn}
\usepackage{bm}
\usepackage[margin=1in]{geometry}
\usepackage[utf8]{inputenc} 
\usepackage[T1]{fontenc}    
\usepackage{hyperref}       
\usepackage{url}            
\usepackage{booktabs}     
\usepackage[b]{esvect}   
\usepackage{amsfonts}       
\usepackage{nicefrac}       
\usepackage{microtype}      
\usepackage{amsmath}
\usepackage{etoolbox}
\usepackage{float}

\usepackage{graphicx}
\usepackage{subfigure}
\usepackage{dcolumn}
\usepackage{bm}
\usepackage{float}
\usepackage{hyperref}

\graphicspath{{Images/}}

\title{Separation of Chaotic Signals by Reservoir Computing}

\author{
	Sanjukta Krishnagopal*, Michelle Girvan, Edward Ott, Brian Hunt\\
	University of Maryland \\
	College Park \\
	Maryland 20742 \\
	*\texttt{sanjukta@umd.edu} \\
}

\begin{document}

\maketitle

\begin{abstract}
We demonstrate the utility of machine learning in the separation of superimposed chaotic signals using a technique called Reservoir Computing. We assume no knowledge of the dynamical equations that produce the signals, and require only training data consisting of finite time samples of the component signals. We test our method on signals that are formed as linear combinations of signals from two Lorenz systems with different parameters. Comparing our nonlinear method with the optimal linear solution to the separation problem, the Wiener filter, we find that our method significantly outperforms the Wiener filter in all the scenarios we study. Furthermore, this difference is particularly striking when the component signals have similar frequency spectra. Indeed, our method works well when the component frequency spectra are indistinguishable - a case where a Wiener filter performs essentially no separation.

\end{abstract}

\section{Introduction}

The problem of extracting a signal from ambient noise has had wide applications in various different fields such as signal processing, weather analysis \cite{BAS04}, medical imaging \cite{CON08}, and cryptography\cite{DOU08}. We consider the related (but potentially more difficult) problem of separating two or more similar signals that are linearly combined. This is a version of the cocktail party problem, i.e., how do people at a cocktail party separate out and focus on the voice of the person they are interested in listening to from a combination of several voices that reaches their ear. In our version of the problem, signals are generated by chaotic processes. If the equations governing these processes are known, the problem can be attacked for example, by using chaos synchronization \cite{BUS07,TSI96,CAR99,ARE06}. We consider instead an approach that relies only on data. Our problem is similar to that of blind source separation for chaotic signals \cite{JIA12,KUR08,SHA13}, but the latter problem typically assumes that multiple linear combinations of the signals are being measured, with the number of independent measurements at least as large as the number of signals. In contrast, our method requires only one linear combination of the signals to be measured after training is complete. For training, we assume that we have finite-time samples of the separate component signals, and our method learns from these samples to separate subsequently measured combination signals into their components. We also note that Wiener obtained an optimal linear solution for the signal separation problem (the `Wiener filter'). In contrast, the technique we use is fundamentally nonlinear, and, as we will show, can significantly outperform the linear technique.

Machine learning techniques have been very successful in a variety of tasks such as image classification, video prediction, voice detection etc. \cite{KRI12,LEG15, KIM17}. Recent work in speech separation includes supervised machine learning techniques such as deep learning \cite{WAN18}, support vector machines \cite{HAN12}, as well as unsupervised methods such as non-negative matrix factorization \cite{SCH06}.
Our approach is based on a recurrent  machine learning architecture originally known as Echo State Networks (originally proposed in the field of machine learning) \cite{JAE01} and Liquid State Machines (originally proposed in the field of computational neuroscience) \cite{MAA02}, but now commonly referred to as Reservoir Computing \cite{LUK09}. Reservoir computing has been applied to several real-world problems such as prediction of chaotic signals \cite{PAT18}, time-series analysis \cite{BUT13}, similarity learning \cite{KRI18}, electrocardiogram classification \cite{ESC14}, short-term weather forecasting \cite{FER08} etc. We expect that, although our demonstrations in this paper are for Reservoir Computing, we expect similar results could be obtained with other types of machine learning using Recurrent Neural Network architectures, like LSTM and Gated Recurrent Units \cite{VLA18}; however, based on results in \cite{VLA18}, we expect that these architectures will have a significantly higher computational cost for training than reservoir computing in the cases we consider. 

We train a Reservoir Computer (RC) as follows. Training data consists of an input time series and a desired output time series. The RC consists of the input layer, the reservoir, and the output layer. The input time series is processed by the input layer as well as the reservoir; the resulting reservoir states are recorded as a vector time series. Then, linear regression is then used to find an output weight matrix (the output layer) that fits the reservoir state to the desired output. The internal parameters of the input layer and the reservoir are not adjusted to the training data, only the output weights are trained.

In this article, we input a linear combination of two chaotic signals to the RC and train it to output an estimate of one of the signals. In the simplest case, which we describe in section \ref{separate}, the ratio between the amplitudes of the signals is known in advance. In section \ref{generalize}, we consider the case in which this ratio is unknown. In this case, we first train a RC to estimate this ratio which we call the mixing fraction, and then train another RC to separate the chaotic signals given the ratio. 
We demonstrate our results on signals from the Lorenz system in the chaotic regime. 

We describe our implementation of reservoir computing in section \ref{reservoir} and review the Lorenz system in section \ref{lorenz_sec}. We compare our results with an approximation to the Wiener filter, computed by estimating the power spectra of the signals using the same training data we use for the RC. If computed from the exact spectra, the Wiener filter is the optimal linear seperator for uncorrelated signals (see Appendix \ref{appendix}). Motivated in part by this comparison, we consider three scenarios is section \ref{separate}: separating signals with different evolution speeds on the same attractor, signals with different Lorenz parameters, and signals with both the parameters and evolution speed perturbed in such a way that the spectra of the signals almost match.  We present our conclusions in section \ref{conclusion}, a brief summary of which are as follows: (1) the RC is a robust and computationally inexpensive chaotic signal separation tool that outperforms the Wiener filter, (2) we use a dual-reservoir computer mechanism to enable separation even when the amplitude ratio of the component signals in the mixed signal is unknown. Here, the first RC estimates a constant parameter, the mixing fraction (or amplitude ratio), whereas the second RC separates signals given an estimated mixing fraction.

\section{Methods}
We consider the problem of estimating two scalar signals $s_1(t)$ and $s_2(t)$ from their weighted sum $u(t) = \beta_1 s_1(t) + \beta_2 s_2(t)$. We normalize $s_1(t)$, $s_2(t)$ and $u(t)$ to have mean 0 and variance 1. 
We assume $s_1(t)$ and $s_2(t)$ to be uncorrelated, in which case our normalization implies, $ \beta_1^2 + \beta_2^2=1$.
Let $\alpha=\beta_1^2$, so that
\begin{equation}
u(t) = \sqrt{\alpha} s_1(t) + \sqrt{1-\alpha} s_2 (t)
\end{equation}
We call $\alpha$ the mixing fraction; more precisely it is the ratio of the variance of the first component of $u(t)$ to the variance of $u(t)$.

In section \ref{separate}, we assume that the value of $\alpha$ is known. In section \ref{generalize}, we consider the case where $\alpha$ is unknown. In both sections, we assume that limited-time samples of $s_1(t)$ and $s_2(t)$ are available, say for $0 \leq t \leq T$, and we use these samples to train the reservoir. Our methods require no knowledge of the processes that generate $s_1(t)$ and $s_2(t)$, but we assume that these processes are stationary enough that the training samples are representative of the components of future instances of u(t). For our numerical experiments, we generate $s_1(t)$ and $s_2(t)$ from the Lorenz family of chaotic dynamical systems (see section \ref{lorenz_sec}). More generally, the same method can be used if $s_2(t)$ is a combination of multiple signals and/or noise.

\subsection{Reservoir Computer}
\label{reservoir}
\begin{figure}[ht!]
	\begin{center}
			\includegraphics[width=0.7\linewidth]{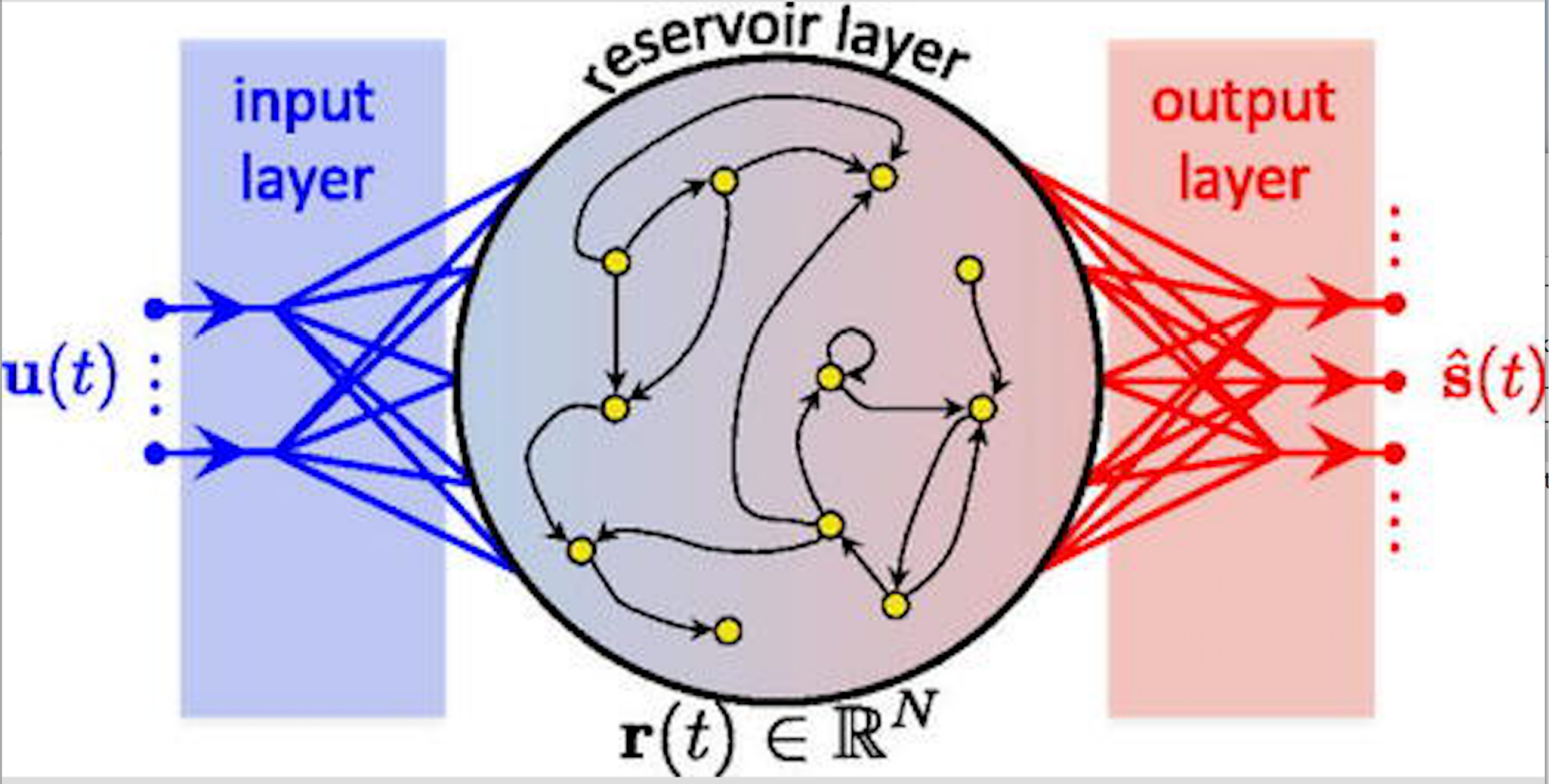}
	\end{center}
	\caption{ Reservoir architecture with input state at time $t$ denoted by $\bm{u(t)}$, reservoir state by $\bm{r(t)}$, and output state by $\bm{\hat{s}(t)}$. The output layer is trained so that $\bm{\hat{s}(t)}$ approximates the desired output signal $\bm{s(t)}$.}
	\label{fig:res}
\end{figure}

There are many variations in implementation; in this paper we adopt the Echo State Network approach of reservoir computing proposed by Jaeger \cite{JAE01}. The reservoir computer has three components (Fig. 1), a linear input layer with $M_i$ scalar input (one for each component of the M-dimensional input signal $\bm{u}(t)$ ), a recurrent, nonlinear reservoir network with $N$ dynamical reservoir nodes driven both by inputs as well as by delayed feedbacks from the other reservoir nodes, and a linear output layer with $M_o$ scalar outputs, as shown in Fig. \ref{fig:res}. We describe a method for general $M_i$ and $M_o$, but in our experiments we will always use $M_i=M_o=1$, with $s(t)$ equal to either $s_1(t)$ or $s_2(t)$, or (in section \ref{generalize}), the constant $\alpha$.

\subsubsection{Input Layer}
The input layer is described by a $N \times M_i$ matrix $W^{\textrm{in}}$ where elements are randomly chosen between to be a uniform distribution between $[-k, k]$, where $k$ is a hyper-parameter to be chosen later.


\subsubsection {Reservoir Layer}
The reservoir can be though of as a dynamical system with state vector $\bm{r}(t)$ at time $t$ given by :
 \begin{equation}
  \bm{r}(t+\Delta t) = (1-a)  \bm{r}(t) + a \tanh{ (\bm{W^{\textrm{in}}}  \bm{u}(t) + \bm{W^{\textrm{res}}}  \bm{r}(t) + b)}
  \end{equation}

The state of the reservoir at time $t$ is given by $\bm{r}(t)$, of size $N$. The notation $tanh(\ldots)$ with a vector argument is defined as the vector whose components are the hyperbolic tangents of the corresponding components of the argument vector. The leakage parameter $a$,  which is bounded between $[0,1]$,  determines the speed at which the input affects or leaks into the reservoir. Both $a$ and the bias magnitude $b$ are hyper-parameters.
The  recurrent  connection  weights  $\bm{W^{\textrm{res}}} \in \mathbb{R}^{N \times N} $ are  initialized  randomly between $-1$ and $1$; then, $\bm{W^{\textrm{res}}}$ is normalized by a multiplication of all its components by a constant chosen so that the spectral radius  (maximal absolute eigenvalue of $\bm{W^{\textrm{res}}}$) $\lambda$, which is another hyper-parameter. A discussion of the effect of spectral radius on performance is presented in \cite{OZT07}. Typically $N$ is much larger than $M_i$, so that the reservoir transforms the input  from the input space into a much higher dimensional reservoir space. The sparsity $sp$ of the reservoir layer is a measure of how sparsely connected the nodes are. Sparsity of zero means all-to-all coupling, and sparsity of one means no connections at all.

\subsubsection{Output Layer}
After running the reservoir (Eq. \ref{res_eqn}) for a transient time period $-100$ to $0$, we form the reservoir state trajectory, $\bm{R}$, is formed by concatenating the reservoir state vectors (the state of all reservoir nodes) at every timestep $\bm{r}(t)$ corresponding to the input $u(t)$ as follows:
\begin{equation}
\bm{R}=[\bm{r}(1), \bm{r}(2) , \ldots , \bm{r}(T)]
\label{res_eqn}
\end{equation}

Thus, $\bm{R}$ is an augmented matrix of size $N \times T$ where $T$ is the training time, i.e., number of time steps for which training data is available. 
During training, the output weights are found by mapping the reservoir state trajectory to the desired output layer representation $\bm{S} = [\bm{s}(1) , \bm{s}(2) , \ldots , \bm{s}(T)]$ over T samples. 
Only the weight matrix $\bm{W^{\textrm{out}}}$ is optimized during training such that  the the mean square error $E(\hat{s} )$ between the output of the reservoir and the target signal $s$ is minimized. The reservoir output $\bm{\hat{s}(t)}$ is obtained through the output weights as follows: 
\begin{equation}
\bm{\hat{s}}(t)=\bm{W^{\textrm{out}}} \bm{r}(t)
\label{s_eqn}
\end{equation} 
 $\bm{W^{\textrm{out}}} \in \mathbb{R}^{ M_o \times N}$ where $M_o$ is the dimensionality of the readout layer.

Ridge Regression (or Thikonov regularization) \cite{WYF08}, is used for fitting, thus making the system robust to overfitting and noise. Ridge regression minimizes squared error while regularizing the Euclidian norm of the weights as follows:
\begin{equation}
  J(\bm{W^{\textrm{out}}})=\eta \|\bm{W^{\textrm{out}}}\|^2 +  \|\bm{W^{\textrm{out}}} \bm{R}- \bm{\hat{s}}\|^2.
\end{equation}
where $\eta$ is a regularization constant, which is also a hyper-parameter.

Once training is done, the RC predicts state variables for times $t \geq T$ through $\bm{W^{\textrm{out}}} $ and Eq. \ref{s_eqn}.

\subsection{Data: Lorenz system}
\label{lorenz_sec}
Our examples are based on the Lorenz equations \cite{LOR63}:

\begin{eqnarray}
\dot{x} &=& \sigma(y-x) \label{eq:X}\\
\dot{y}  &=& -xz + \rho x -y \label{eq:Y}\\
\dot{z}  &=& xy - \beta z \label{eq:Z}
\label{lorenz}
\end{eqnarray}

\noindent 
The Lorenz attractor is a strange attractor that arises in these system of equations. The Lorenz system is known to be chaotic for the parameter values:
$\sigma = 10$,
$\rho = 28$ and
$\beta = 8/3$ \cite{TUC02}, and appears to be chaotic for other parameter values we use in this article. We generate trajectories using the $4^{th}$ order Runge-Kutta method with step size 0.01, and we sample them every five steps, i.e., 0.05 time units.

\section{Results: Generalization Separation with Known Mixing Fraction}
\label{separate}

We calculate the error between the trained reservoir output $\hat{s}(t)$ and the desired signal $s_1(t)$ as follows. The mean square reservoir error $E_R$ is:
\begin{equation}
E_R={\frac{<({s_1} - {\hat{s}})^2> }{\min_\zeta<({s_1} - \zeta {u})^2>}} 
\end{equation}
where $s_1 \in [x_1, y_1, z_1]$, i.e., one of the components of the signal (typically we use $s_1=x_1$). $\zeta$ scales the input so that the denominator indicates the root mean square error in the absence of any processing by the reservoir.

The Wiener filter \cite{WIE49} is known to be the best possible  \textit{linear} solution to our signal separation problem, and is also commonly used in tasks such as signal denoising, image deblurring etc. A detailed explanation of the Wiener filter is given in Appendix \ref{appendix}. In this article, we always compare the RC with a Wiener filter, which uses a window of $500$ for estimating the spectrum.
 The mean square Wiener error measure $E_W$ is:
\begin{equation}
E_W={\frac{<({s_1} -{\hat{s}}_w)^2> }{\min_\zeta<({s_1} - \zeta {u})^2>}} 
\end{equation}
where $\hat{s}_w$ is the output of the Wiener filter. 

\subsection{Separating Lorenz signals with different parameters}
\label{diffparams}
In this section, the reservoir input consists of a combination of two x-component signals from Lorenz systems with different parameter values. The signal $x_1$ has parameters $ \bm{p}_1 = \{\sigma = 10,
\rho = 28,
\beta = 8/3\}$. The signal $x_2$ has parameters $\bm{p}_2= 1.20 \times \bm{p}_1$. 

\begin{figure}[ht!]
	\centering
	\includegraphics[width=.9\linewidth]{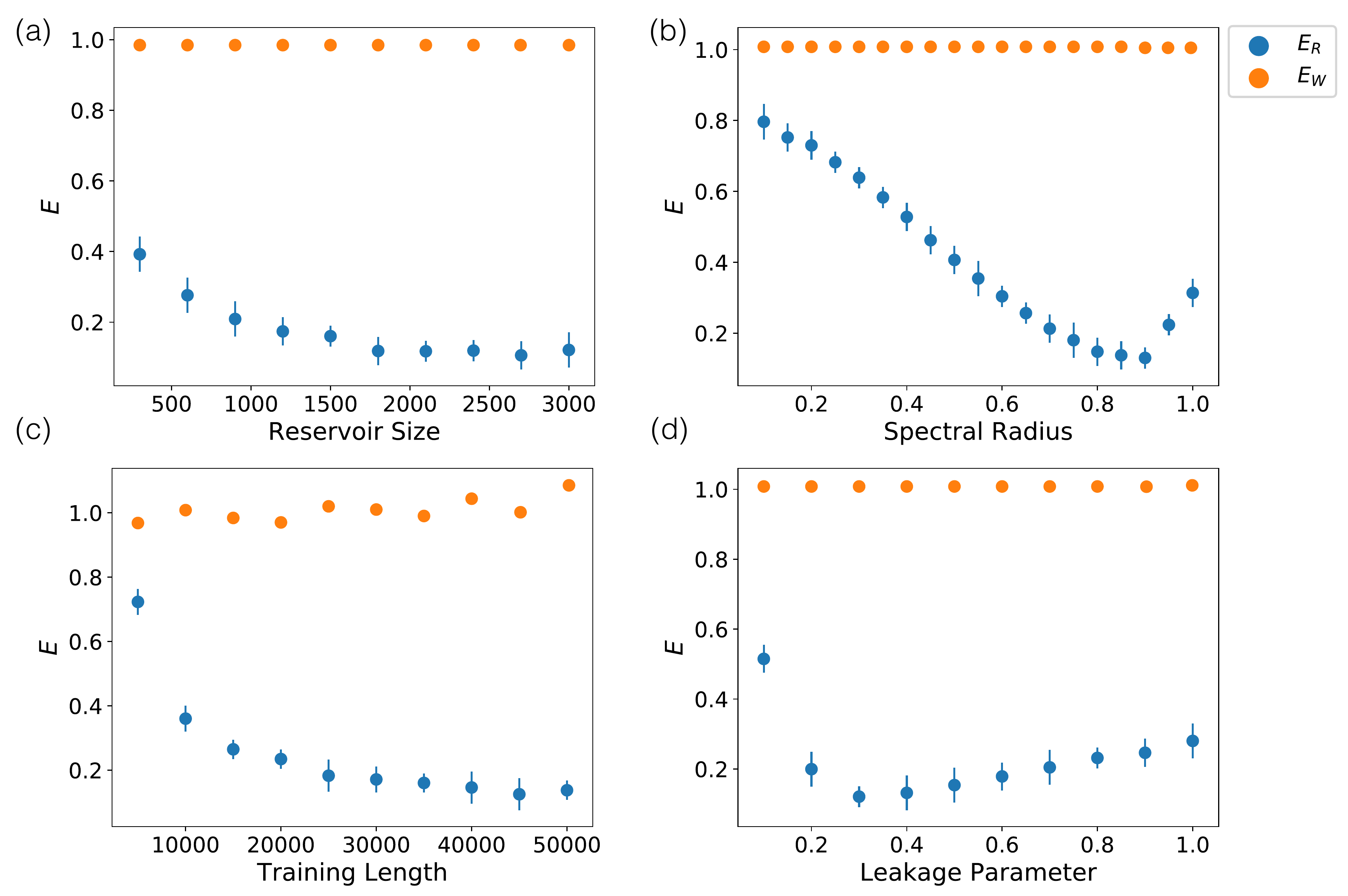}
	\caption{ Reservoir error $E_R$ and Wiener error $E_W$ over a test length of 5000 timesteps for (a) varying reservoir size $N$, (b) spectral radius $\lambda$, (c) training length and (d) leakage parameter $a$. All parameter values except the parameter being varying: $\Delta t=0.05$, sparsity $sp=0.95$, $N=2000, \lambda =0.9, a= 0.3, k=0.13$ training length $=50000, \alpha = 0.5$, $\bm{p}_2=1.2 \bm{p}_1$.}
	\label{params}
\end{figure}

Several parameters related to the reservoir system must be chosen for running experiments. These parameters include reservoir size $N$, spectral radius $\gamma$, leakage parameter $a$, and the length of training signal. In Fig. \ref{params}, we vary each of these parameters individually while keeping the others constant in order to identify an appropriate set of values.

Fig. \ref{params} shows the performance of the reservoir in separating the $x$ component of the $s_1$ trajectory for varying parameters given $\alpha = 1/2$. The reservoir errors are averaged over 10 random initializations of the reservoir. The error bars denote standard error, i.e., the standard deviation across $l$ random initializations over $\sqrt{l}$. We observe a downward trend in error with increasing reservoir size, which seems to saturate at about $2000$ nodes. Panel (a) shows that a reservoir size of $N=2000$ gives a reasonable trade-off between performance and computational efficiency. Panel (b) shows that $\lambda=0.9$ seems to result in the best separation, with all other parameters constant. 
As seen in panels (c,d), the optimal leakage parameter $a$ and training length are $0.3, 50000$ respectively. Other parameters chosen are input strength normalization $k=0.13$ such that the input does not saturate the tanh curve. These parameter values are used for the rest of this paper unless mentioned otherwise. 

 \begin{figure}[ht!]
	\centering
	\includegraphics[width=.9\linewidth]{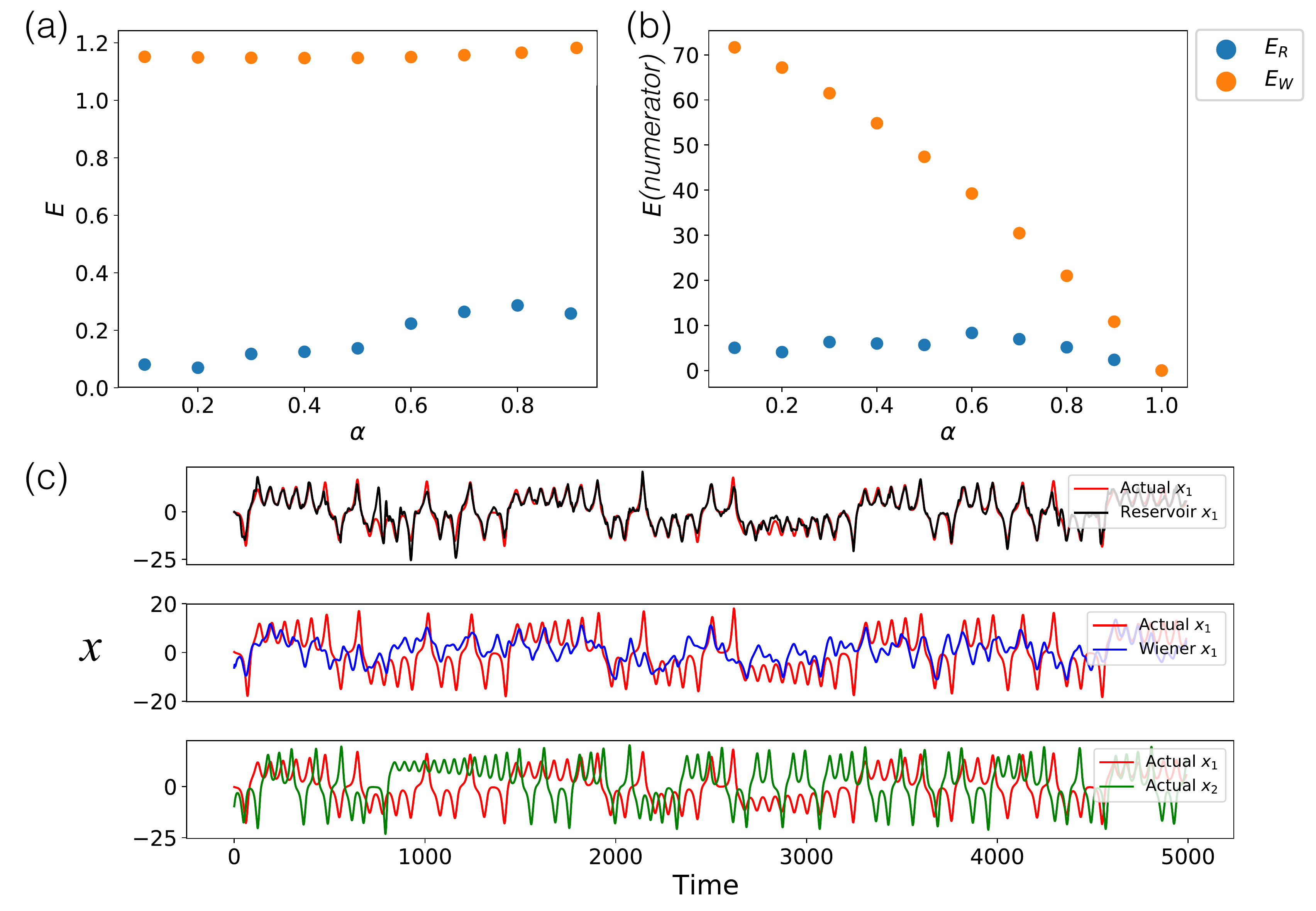}
	\caption{ (a) Reservoir (blue) and Wiener (orange) error over a test length of 5000 timesteps across mixing fraction $\alpha$ for two Lorenz signals with the same speed but parameters $\bm{p}_1 = 1.2 \times \bm{p}_2$. (b) shows the numerator only of the error measures $E_R, E_W$. (c) For a mixing fraction $\alpha= 0.5$, (top) time-series plot of actual $x_1$ and reservoir predicted output $\hat{s}$, (middle) time-series plot of actual $x_1$ and Wiener predicted output $\hat{s}_w$, (bottom) actual $x_1$ and $x_2$.  $E_R=0.15,E_W=1.14$. $\Delta t=0.05$, sparsity $sp=0.95$, $N=2000, \lambda =0.9, a= 0.3, k=0.13$, training length $=50000$, and for (c), $\alpha = 0.5$.}
	\label{diff_params}
\end{figure}

Fig. \ref{diff_params} (a) presents the performance of the reservoir as a function of the mixing fraction. We observe that the reservoir computer consistently outperforms the Wiener filter. However, as $\alpha$ increases, the error increases as well. This can be attributed to the denominator in the the error measure tending to zero as alpha tends to one. 
Fig. \ref{diff_params} (c) shows the estimated separated $x_1$ trajectory along with the actual $x_1$ in the testing phase, and the actual $x_1, x_2$ trajectories. The top panel demonstrates that the RC prediction does indeed match the actual $x_1$ accurately. The middle panel demonstrates that the best linear filter, the Wiener filter, is comparitively worse than the RC at estimating the chaotic signal $x_1$. 

\subsection{Separating Lorenz signals with different speed}
\label{diffspeed}

In this section, the reservoir input consists of a combination of two signals from the $x$ component of the Lorenz system with the same parameter values $\bm{p}_2=\bm{p}_1$, but with different speeds, i.e., the right hand side of Eq. \ref{lorenz} for the Lorenz system corresponding to $x_1$ is multiplied by a speed fraction (ratio of speeds) $\eta$. The reservoir parameters remain the same as those found in sec. \ref{diffparams}.

 \begin{figure}[ht!]
	\centering
	\includegraphics[width=.9\linewidth]{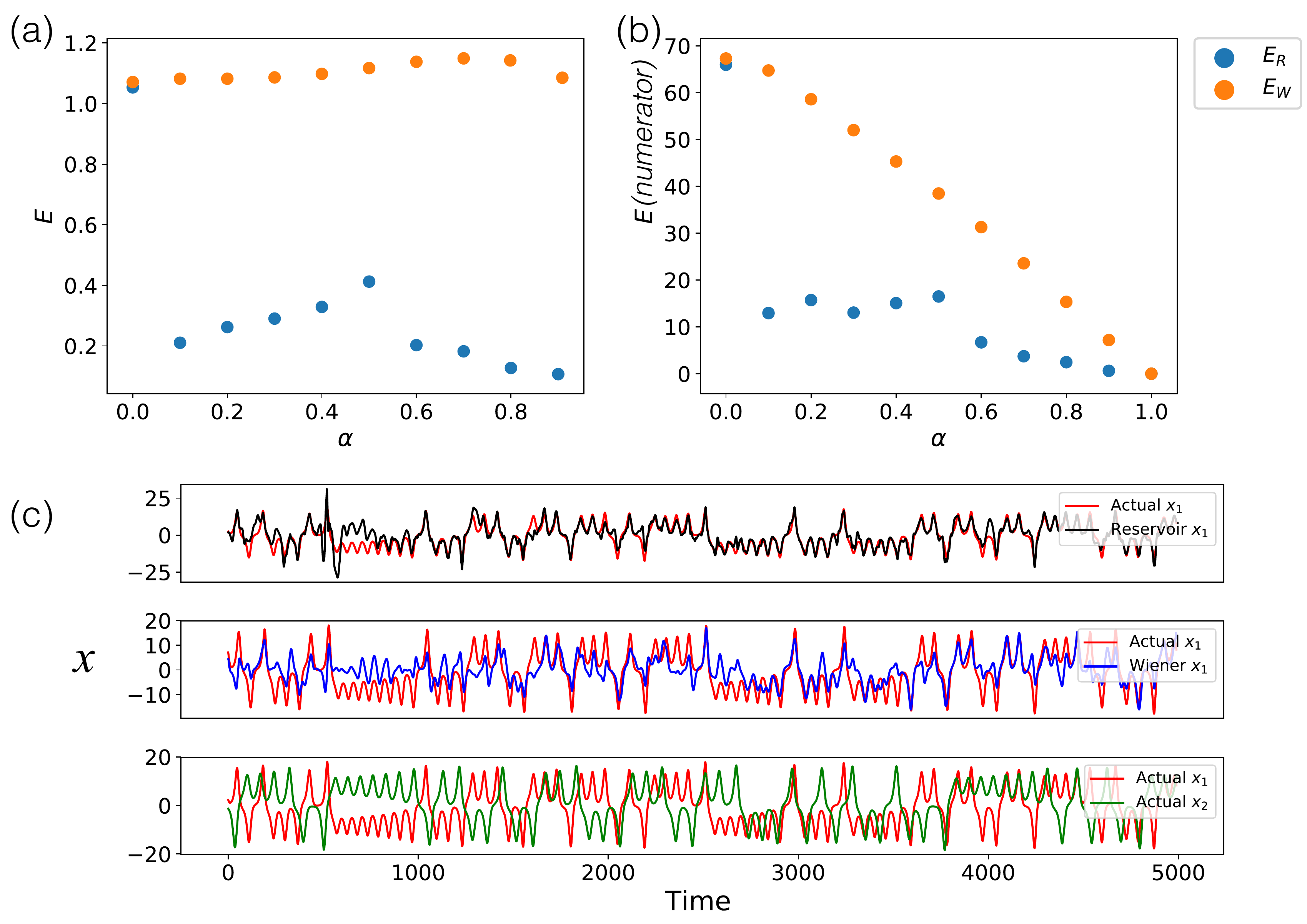}
	\caption{ (a) Reservoir (blue) and Wiener (orange) error over a test length of 5000 timesteps across mixing fraction $\alpha$ for two Lorenz signals with the same parameters but speeds $\eta_1 = 1.2 \times \eta_2$. (b) shows the numerator only of the error measures $E_R, E_W$. (c) For a mixing fraction $\alpha= 0.5$, (top) time-series plot of actual $x_1$ and reservoir predicted output $\hat{s}$, (middle) time-series plot of actual $x_1$ and Wiener predicted output $\hat{s}_w$, (bottom) actual $x_1$ and $x_2$. $E_R=0.51,E_W=1.02$. $\Delta t=0.05$, sparsity $sp=0.95$, $N=2000, \lambda =0.9, a= 0.3, k=0.13$, training length $=50000$, and for (c) $ \alpha = 0.5$.}
	\label{diff_speed}
\end{figure}

Fig. \ref{diff_speed} (a) presents the performance of the reservoir computer as a function of the mixing fraction for $\eta_1=1.2 \eta_2$ respectively, where $\eta_i$ is the speed of the $i^{th}$ signal. The error seems to have an increasing trend with $\alpha$, as in the case with different parameters as in section \ref{diffparams}. This can be attributed to the denominator in the error measure tending to zero as alpha tends to one. The non-normalized error is plotted in (b), and here we can see that as $\alpha$ tends to one, error tends to zero, as it should. The RC consistently outperforms the Wiener filter.  As seen in Fig. \ref{diff_speed} (c), the Wiener prediction is much poorer than the reservoir prediction of the chaotic signal $x_1$.

\subsection{Separating Lorenz signals with matched power spectras}
\label{samespectra}
\begin{figure}[ht!]
	\centering
	\includegraphics[width=.9\linewidth]{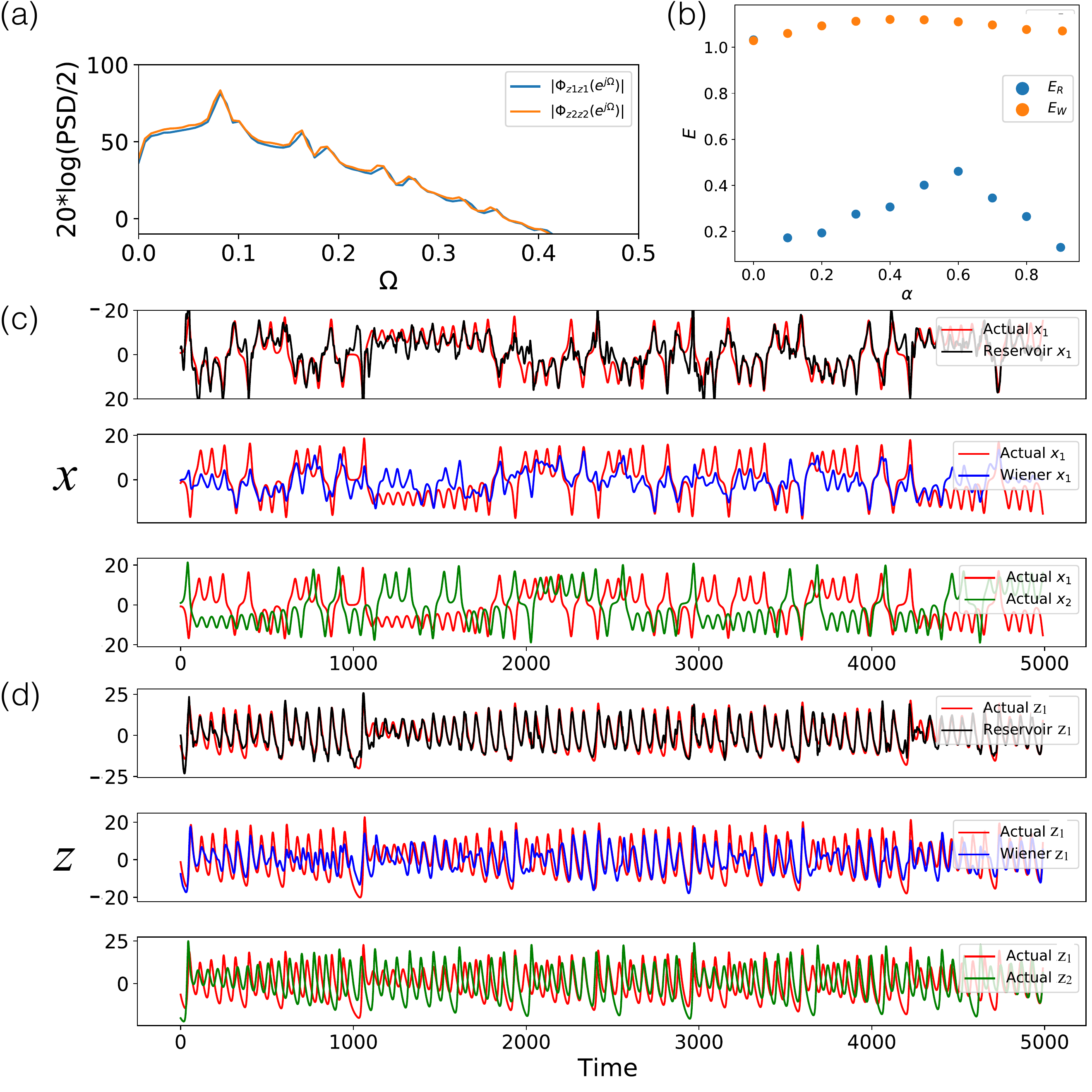}
	\caption{(a) Power spectral density of $z_1,z_2$ across frequency $\Omega$, on a scaled log-plot to demonstrate a clear match in PSD respectively. (b) shows Reservoir and Wiener error ($E_R,E_W$) for separation of spectra matched Lorenz $x$ signals as a function of $\alpha$.  (c,d) For a mixing fraction $\alpha= 0.5$, (top) time-series plot of actual $x_1,z_1$ respectively and reservoir predicted outputs, (bottom) actual inputs $x_1,x_2$ and $z_1,z_2$ respectively. for (c), $E_R=0.48,E_W=1.02$ and for (d), $E_R=0.26,E_W=1.14$.  $t=0.05$, sparsity $sp=0.95$, $N=2000, \lambda =0.9, a= 0.3, k=0.13$ training length $=50000, \alpha = 0.5, \eta_1=1.1 \times \eta_2, \bm{p}_2= 1.1 \times \bm{p}_1$.}
	\label{match}
\end{figure}

Linear signal denoising/separation methods are based on difference in spectra. Here we study the $z$ component of two signals, $z_1, z_2$ with $\bm{p}_2=1.1 \times \bm{p}_1$ and $\eta_2 = 0.9 \times \eta_1$, where $ \bm{p}_1 = \{\sigma = 10,
\rho = 28,
\beta = 8/3\}$, and $\eta_1 =1$. Fig. \ref{match} (a) shows a log plot of the Power Spectral Density (PSD) $\phi_{z1z1}, \phi_{z2z2}$ of the two signals $z_1, z_2$ respectively as a function of frequency $\Omega$. The PSD of the $z$ component of the Lorenz system has a distinct peak (unlike the $x,y$ components), followed by a much smaller peak. For this reason we plot the PSD of the $z$ signals to demonstrate spectra matching even though we separate the corresponding $x$ signals (which may conceivably be a harder problem to solve since the $x$ switches between positive and negative sides chaotically, and the RC has to learn when to switch correctly). We observe, in panel (a) that the peaks do indeed line up and the spectra match fairly well. Fig. \ref{match}(b) plots the reservoir and Wiener errors as a function of $\alpha$ for the spectra matched case. For a case where the spectra of the two individuals are indistinguishable, a spectra-based filter such as the Wiener filter will be unable to separate the signals. Fig. \ref{match}(c,d) show the reservoir predictions for the $x$ and $z$ component of the Lorenz system respectively for $\alpha=0.5$. We observe, that the RC is indeed, able to separate the signals, even when their spectra match, unlike other state-of-the-art spectra-based signal separation methods like the Wiener filter (see Appendix \ref{appendix}). 

\section{Generalization Separation with Unknown Mixing Fraction}
\label{generalize}
Often, as in the case of the cocktail party problem, the ratio of amplitudes in a mixed signal may not be known. In this section, we describe a methodology to separate signals without knowledge of the mixing fraction $\alpha$. We found that training a single reservoir computer to separate signals for a wide range of $\alpha$ values was unsuccessful. So here we present a two step method: First, train a single RC to identify the mixing fraction (output $\alpha$ given a mixed signal). As before, we assume that we have access to training data for the individual signals. Using this kind of data along with known mixing fractions, we can first train a single RC to identify the mixing fraction of a mixed signal with an unknown mixing fraction. We then train a second RC to separate the signals using this estimated $\alpha$ as in section \ref{separate}.

\subsection{Estimation of the Mixing Fraction Parameter}
In training, the first RC is given mixed signals over a range of discrete values of $\alpha$ (0 to 1 in intervals of 0.05) and trained to output a constant $\alpha$ value. In testing, the constant predicted $\alpha$ are averaged over the testing time. However, learning a fixed value from a chaotic signal is non-trivial. We find that the trained RC always has a tendency to predict $\alpha$ closer to mean $\alpha$ (0.5) when fit on both the training and the test data, i.e., overestimate small actual $\alpha$ and underestimate large actual $\alpha$. To correct for this tendency, we use the training dataset to construct a mapping from the predicted $\alpha$ to the desired $\alpha$ via a fit to a third order polynomial function. We then apply this same function to the reservoir-predicted $\alpha$ value(s) for the test data, in order to obtain corrected values.

\begin{figure}[ht!]
	\centering
	\includegraphics[width=.8\linewidth]{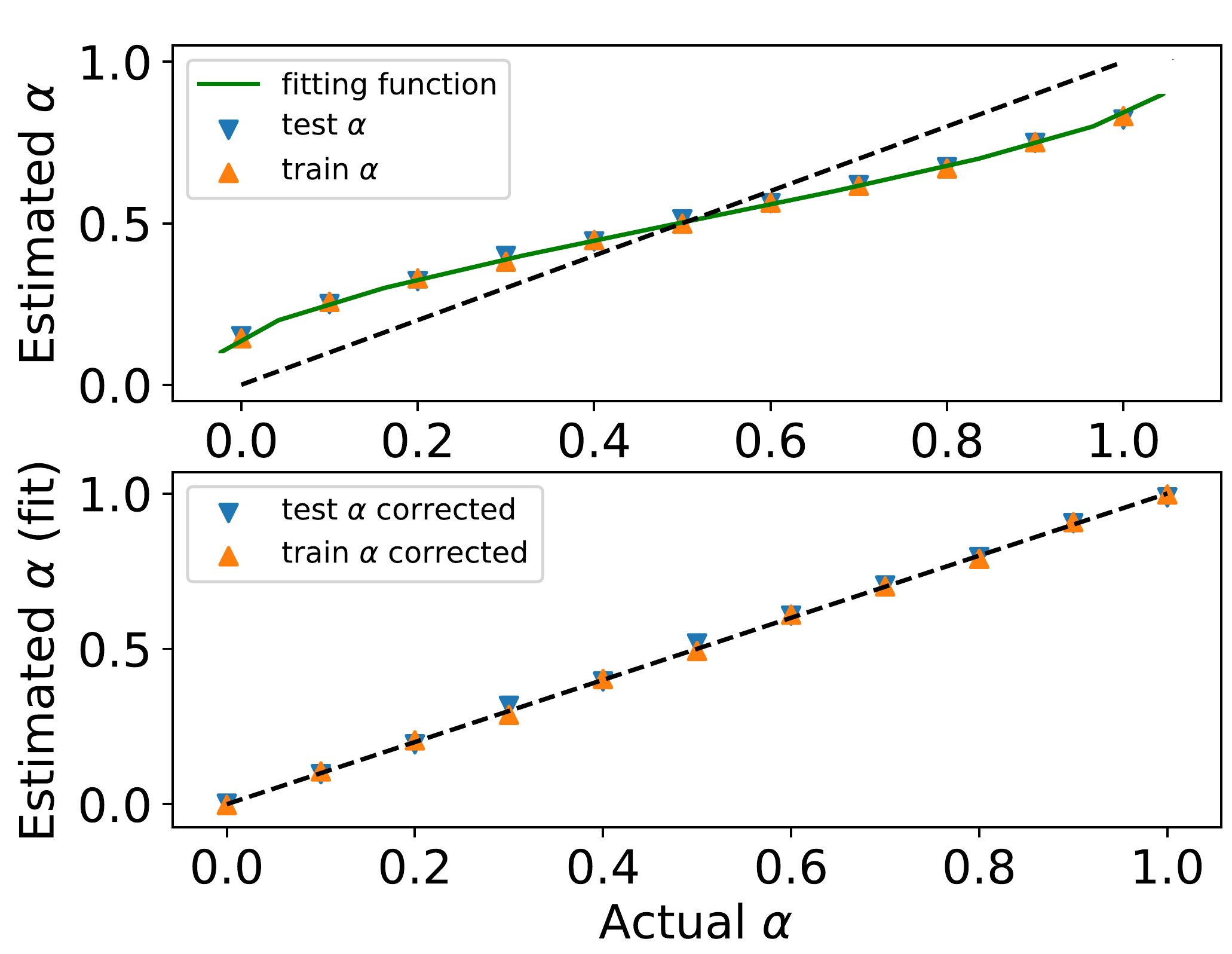}
	\caption{ (top) Plot of the actual $\alpha$ used in the mixed signal vs the reservoir predicted $\alpha$ for both the training (orange) and test (blue) dataset. The green curve is the third degree polynomial fitting function between 0 and 1. The black dashed line represents the diagonal for reference. (bottom) The corrected test and train RC-estimated mixing fractions (after fitting to the third degree polynomial). The Lorenz signals being separated have the following characteristics $\bm{p}_2=1.2 \bm{p}_1, \eta_1=\eta_2$. Here $\Delta t=0.05$, sparsity $sp=0.99$, $N=1000, \lambda =0.9, a= 0.3, k=0.13$ such that input signal has unit standard deviation, training and testing length $=50000$ for each training $\alpha =[0,0.1,0.2, \ldots, 0.9,1]$.}
	\label{gen}
\end{figure}

Fig. \ref{gen} illustrates the performance of our approach for separating a mixed signal of x-components of two Lorenz systems with parameter values that differ by 20\%. Fig. \ref{gen} (top) shows the reservoir-estimated $\alpha$ vs the actual $\alpha$ and Fig. \ref{gen} (bottom) shows the corrected $\alpha$ estimate (calculated using a third-order polynomial fit to the points in \ref{gen} (top)). After making the correction, our method accurately predicts $\alpha$ for both the training and test data. 


Once the RC predicts an estimate of the mixing fraction, a second RC can be trained on that value of mixing fraction to separate chaotic Lorenz signals. 

We note also that similar to this method for estimating the mixing fraction parameter, RC along with output correction can be used for parameter estimation from data in other systems as well.

\subsection{Interpolating between trained reservoir computers}

Often, the RC may have computational and consequently training constraints, i.e., it may not be feasible to train a RC on each value of predicted $\alpha$ obtained. For instance, RCs in hardware have training constraints. In such cases, RCs can be pre-trained on discrete values of $\alpha$. Any intermediate predicted $\alpha$ estimate obtained can then be used for separation by interpolating between the two nearest trained $\alpha$ RCs. Here, interpolating between RCs means interpolating between their trained output weights. What spacing of $\alpha$ is appropriate for training? A mixed signal with predicted $\alpha=q$ can  be separated by using the following output weight matrix $\bm{W^{out}_q}$ if $q$ is in between two discrete trained values of $\alpha$ ($q_-$ and $q_+$).
\begin{equation}
\bm{W^{out}_q} = \frac{q - q_-}{q_+-q_-} \bm{W^{out}_{q_+}} + \frac{q_+ - q}{q_+-q_-}  \bm{W^{out}_{q_-}}
\label{ave}
\end{equation}

\begin{figure}[ht!]
	\centering
	\includegraphics[width=.9\linewidth]{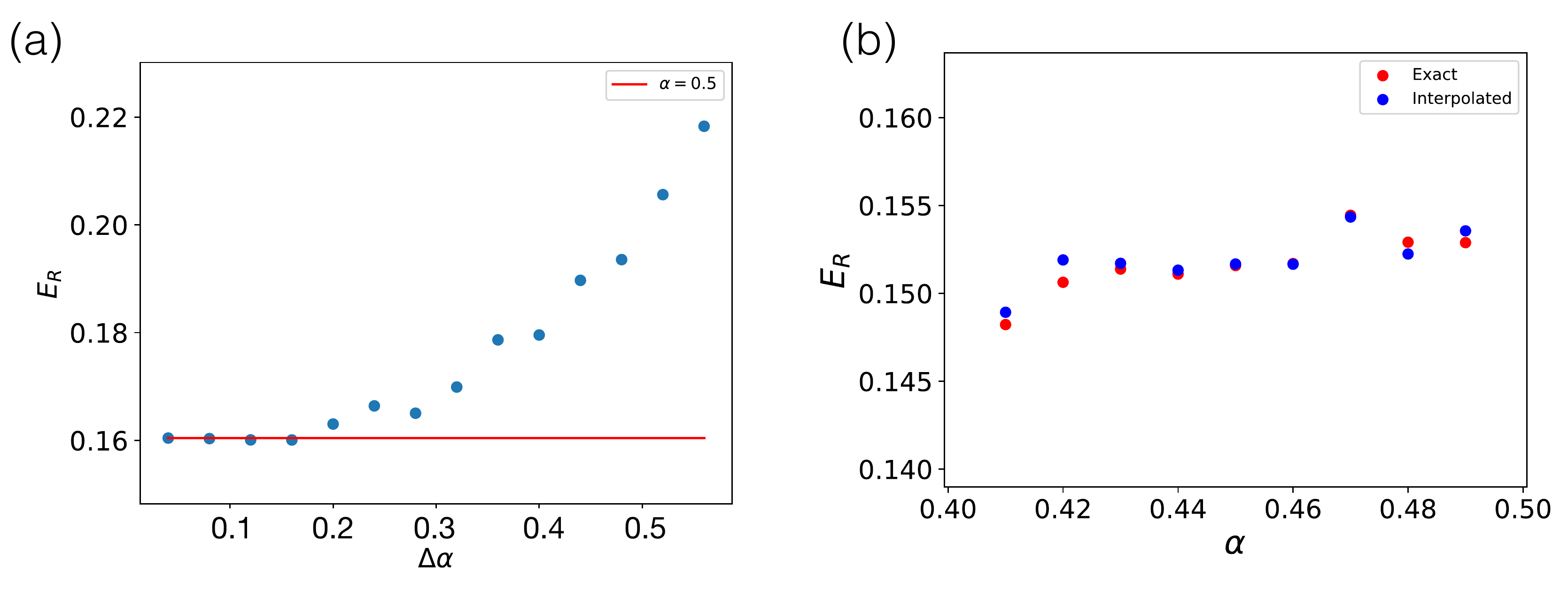}
	\caption{ Reservoir errors on separating Lorenz signals with $\alpha=0.5$. (red) RC trained on $\alpha=0.5$, (blue) average of RCs trained on $\alpha=0.5 \pm \Delta \alpha/2$ (b) Errors in interpolation between predictions by independent RCs trained on $\alpha =[0.4,0.5]$ (blue) and RC trained on the exact $0.4<\alpha<0.5$ (orange); training length $=50000$, training length =$5000$. $\bm{p}_2=1.2 \bm{p}_1, \eta_2= \eta_1$, $\Delta t=0.05$, sparsity $sp=0.95$, $N=1000, \lambda =0.9, a= 0.3, k=0.13$}
	\label{interpolation}
\end{figure}

In Fig. \ref{interpolation}, (a), we plot the reservoir error $E_R$ for $\alpha=q=0.5$ in orange, and for the average of the $\bm{W^{out}}$ matrices of reservoirs trained on $\alpha \pm \Delta \alpha/2$. We notice that interpolating between discrete RCs with a spacing of $\Delta \alpha = 0.1$ is well within the range of negligible error compared to training on individual predicted $\alpha$s. Hence, one only needs to train on $\alpha$s in intervals of 0.1 ($\alpha= [0,0.1,0.2,\ldots,0.9,1.0])$. Fig. \ref{interpolation}, (a) shows the reservoir error for RCs trained individually on $\alpha =[0.40,0.41,0.42...0.49,0.50]$ compared with the reservoir error obtained by averaging the $\bm{W^{out}}$ matrices of RCs trained on $\alpha=0.4, 0.5$ appropriately as in Eq. \ref{ave}. We observe that the results practically coincide, and that the method is fairly robust to errors in $\alpha$ prediction. 
Successful interpolation between RCs trained on discrete mixing fractions drastically reduces the training time and computation without compromising on quality of signal separation. In fact, we only need to train on 11 distinct mixing fractions to be able to separate Lorenz signals mixed in any proportion.  Thus we are able to generalize our chaotic signal separation technique to cases where the mixing fraction is unknown.

\section{Conclusion and Future Work}
\label{conclusion}

In this article, we used a type of machine learning called reservoir computing for separation of chaotic signals. We demonstrated the ability to separate signals for several cases where the two signals are obtained from Lorenz systems with different parameters. We compared out results with the Wiener filter, which is the optimal linear filter and whose coefficients can be computed from the power spectra of the signals (see Appendix \ref{appendix}). Spectra-based methods, naturally, perform poorly at signal separation if the spectra of the signals to be separated are indistinguishable. By contrast, the RC performs reasonably well even when the two signals that are mixed have very similar spectra. Our results were significantly better than the Wiener filter calibrated from the same training data for all the scenarios we considered.

Often, in signal separation applications such as the cocktail party problem, the mixing fraction (amplitude ratio) of the signals to be separated is unknown. We introduce a RC-based way to separate signals with an unknown parameter, in our case the mixing fraction. The first step of this generalized method is estimating a mixing fraction for a given signal. Estimation of a constant valued parameter from a temporal signal is a problem of broad interest, with applications such as weather prediction, predicting parameters of flow, equation modeling etc. We find that after time-averaging its trained output, the RC tends to skew the estimated parameter towards the mean of the parameter values used during training. By fitting a mapping function that corrects the averaged output in the training data, we are also able to approximately correct the test output. This method of introducing an additional non-linear 'correction' to the learned output weights may be useful for predicting constant outputs in other dynamic machine learning systems. 

In some cases, training on a wide range of mixing fractions may not be possible, due to the need for quick separation of signals and limited training capacity (e.g. in hardware applications). The need for quick separation of signals and limited training capacity may limit the ability of the RC to train on arbitrary mixing fractions (for instance in hardware applications). Hence, we study the robustness of the RC, and the ability to use interpolated RCs pre-trained at discrete mixing fractions. We demonstrate that the RCs need only be trained at a coarse grid of mixing fractions in order to accurately separate Lorenz signals with arbitrary mixing fractions. Our results are robust to errors in the prediction of mixing fraction. Hence, in situations where computational resources are constrained, RCs can be trained on discrete mixing fractions, and interpolation between these RCs can be used to accurately separate chaotic signals for intermediate values of mixing fraction. 

Here we have demonstrated the ability of reservoir computing to act as an efficient and robust method for separation of chaotic signals. The dynamical properties of the reservoir make it a prime candidate for further exploration of chaotic signal processing. An interesting future direction might be to study alternate RC architectures such as parallel RCs for more complex signal extraction problems. 

\section*{ Acknowledgment}
This research was supported through a DoD contract under the Laboratory of Telecommunication Sciences Partnership with the University of Maryland as well as NSF award DGE-1632976.

\nocite{*}
\bibliographystyle{plain}
\bibliography{ms}

\appendix\section{Appendix: Wiener Filter}
\label{appendix}

The Wiener filter \cite{WIE49} was designed to separate a signal from noise, but it can be used to (imperfectly) separate any two signals with different power spectra.  For simplicity, we formulate the filter here in continuous time, as a linear noncausal filter with infinite impulse response (IIR) for a scalar signal. Later we describe how we compute the filter in practice, in discrete time with finite impulse response.

Let ${u}(t)$ be the combined signal -- the input to the filter -- and let ${s}(t)$ be the component signal that is the desired output of the filter.
A noncausal IIR filter can be written as a convolution
\begin{equation}\label{eq:a1}
h*{u}(t) = \int_{-\infty}^\infty h(\tau) {u}(t-\tau) d\tau
\end{equation}
where $h(\tau)$ is the impulse response function of the filter.  (We assume that $|h(\tau)|$ has finite integral; then if ${u}(t)$ is bounded, so is $h*{u}(t)$.)  Of all such filters, the Wiener filter is the one that minimizes the mean-square error $\langle (h*{u}(t) - {s}(t))^2 \rangle$ between the filter output and  the desired output.


Similar to linear least-squares in finite dimensions, the minimizing function $h_W$ can be related to the auto-covariance and cross-covariance functions
\begin{equation}\label{eq:a2}
C_{uu}(\tau) = \langle {u}(t-\tau) {u}(t) \rangle, \qquad
C_{us}(\tau) = \langle {u}(t-\tau) {s}(t) \rangle.
\end{equation}
Specifically, setting the first variation (\textit{i.e.}, the first derivative in the calculus of variations) of the mean-square error equal to zero yields
\begin{equation}\label{eq:a3}
h_W*C_{uu}(\tau) = C_{us}(\tau).
\end{equation}
Equation~(\ref{eq:a3}) can be solved in the frequency domain, where convolution becomes multiplication.  Let $H_W$, $P_{uu}$, and $P_{us}$ be the Fourier transforms of $h$, $C_{uu}$, and $C_{us}$ respectively; then $H(\omega) P_{uu}(\omega) = P_{us}(\omega)$, and
\begin{equation}\label{eq:a4}
H_W(\omega) = \frac{P_{us}(\omega)}{P_{uu}(\omega)}.
\end{equation}

We interpret equation~(\ref{eq:a4}) based on the fact that $P_{uu}(\omega)$ and $P_{us}(\omega)$ are respectively the power spectral density of $u(t)$ and the cross spectral density of $s(t)$ and $u(t)$, by an appropriate version of the Wiener-Khinchin theorem.  In both Wiener's application and our application, the component signals $s(t)$ and $s=u(t) - s(t)$ are uncorrelated, in the sense that their cross-covariance is identically zero.  In this case, $C_{us}(\tau) = C_{uu}(\tau)$ and $P_{us}(\omega) = P_{uu}(\omega)$.  Thus, the transfer function $H_W(\omega)$ of the Wiener filter at frequency $\omega$ is the fraction of the power of $u(t)$ at that frequency attributable to the component signal $s(t)$.

In practice, we sample the signals $s(t)$ and $u(t)$ at discrete intervals of $\Delta t = 0.05$.  We estimate their spectral densities using the Welch method, averaging the estimates on overlapping segments of length $500$ samples, using the discrete Fourier transform (DFT) and the Hann window on each segment.  We then apply the inverse DFT to the resulting discrete estimate of $H_W$, yielding an impulse-response vector ${h}_w$ of length $500$.  We convolve ${h}_w$ with the sampled signal to compute the Wiener filter.

\end{document}